\documentclass[aps,twocolumn]{revtex4}
\usepackage{epsf}
 
\begin{document}
      \title{Influence of the pair coherence on the charge 
             tunneling\\ through a quantum dot connected 
	     to a superconducting lead}

     \author{T.\ Doma\'nski, A.\ Donabidowicz and K.I.\ Wysoki\'nski}
     
\affiliation{
             Institute of Physics, 
	     M.\ Curie Sk\l odowska University,
             20-031 Lublin, Poland} 
      \date{\today}

\begin{abstract} 
We analyze the charge transport through a single level quantum 
dot coupled to a normal (N) and superconducting (S) leads where 
the electron pairs exist either as the coherent (for temperatures 
below $T_{c}$) or incoherent objects (in a region $T_{c} < T <T^{*}$). 
This situation can be achieved in practice if one uses the high 
$T_{c}$ superconducting material where various precursor effects 
have been observed upon approaching $T_{c}$ from above. 
Without restricting to any particular microscopic mechanism 
we investigate some qualitative changes of the nonequilibrium 
charge current caused by the electron pair coherence.
\end{abstract}

\maketitle

\section{Introduction}
It is a firm experimental fact that the phase transition from 
normal to superconducting states of the underdoped high $T_{c}$ 
materials is accompanied by appearance of a pseudogap \cite{review}.
While approaching the critical temperature $T_{c}$ from above 
the single particle states are gradually depleted in a certain 
energy region $ \left| \omega \right| \!<\! \Delta_{pg}$ 
around the Fermi level. This phenomenon is often interpreted 
theoretically as a precursor of the true superconducting gap 
which is usually predicted by the BCS-like treatments for 
temperatures $T<T_{c}$. On a microscopic basis the pseudogap
can be assigned to presence of the electron pairs. Above $T_{c}$  
their long-range coherence is destroyed by the strong quantum 
fluctuations arising either from the reduced dimensionality, 
or due to close neighborhood of the Mott insulating state 
or because of a competition  with various kinds of ordering 
in the system. The incoherent electron pairs have been 
unambiguously detected experimentally above $T_{c}$ in 
measurements of the large Nernst coefficient \cite{Wang-05} 
and in observation of the Berezinski-Kosterlitz-Thouless phase 
fluctuations \cite{Corson-99}. There is however a great 
amount of controversy regarding the temperature extent where  
the incoherent pairs eventually exist. According to available 
experimental data their existence is established for at least 
a dozen Kelvin above $T_{c}$ but such region can perhaps 
spread over a much wider regime up to $T^{*}$ at which the 
pseudogap finally closes. 

Various tunneling techniques have been used for a long time 
as a useful tool for probing the single particle spectra of 
the correlated systems. Recent technological progress of the 
spectroscopic methods such as the STM \cite{Fisher-06}, the 
{\bf k}-resolved photoemission (ARPES) \cite{Damascelli-03}, 
the Andreev-type techniques \cite{Deutscher-05} and the Fourier 
transformed scanning tunneling spectroscopy \cite{McElroy-05} 
allow for a precise measurements of the energy, momentum and 
space dependent  density of states. In the present work we 
propose to consider the N-S junction with the quantum dot 
located in between. In this setup one would be able not only 
to detect the (pseudo)gap in the single particle excitation 
spectrum but, due to activation of additional transport channels 
(mainly the Andreev reflections), there would be a possibility 
to analyze the electron pair coherence. 

Our discussion here is not limited to any particular microscopic 
model describing the electron pair formation and their eventual 
coherence. On rather general grounds we investigate the proximity 
effect which leads to a particle-hole mixing at small energies 
$|\omega| \leq \Delta$ and furthermore we analyze its influence 
on the effective charge transport through the quantum dot. In 
section II we briefly introduce the problem, in particular 
explaining how we treat the coherent and incoherent electron 
pairs. Next, we show the QD spectrum without any correlations 
(section III) and for the limit of very strong correlations 
(section IV). The main part of our study is in the section V, 
where we discuss the differential conductance as a function 
of the applied bias $V$ and temperature $T$, both below and 
above $T_{c}$.

\section{Formulation of the problem}

For a description of the quantum dot (QD) coupled to a normal 
($N$) and superconducting ($S$) leads we consider the single 
impurity Anderson model
\begin{eqnarray}
\hat{H} & = & \hat{H}_{N} + \hat{H}_{S} +
\sum_{\sigma} \epsilon_{d} \hat{d}^{\dagger}_{\sigma}
\hat{d}_{\sigma} +  U \; \hat{n}_{d \uparrow} 
\hat{n}_{d \uparrow} \nonumber \\
& + & \sum_{{\bf k},\sigma} \sum_{\beta=\{ N,S \} } 
\left[ V_{{\bf k} \beta} \; \hat{d}_{\sigma}^{\dagger} 
\hat{c}_{{\bf k} \beta \sigma} + \mbox{h.c.} \right] .
\label{model}
\end{eqnarray}
Operators $d_{\sigma}$ ($d_{\sigma}^{\dagger}$) annihilate 
(create) the QD electrons whose energy is $\varepsilon_{d}$. 
The Coulomb potential $U\!>\!0$ describes a repulsion between 
electrons of opposite spin $\sigma$ and the hybridization 
$V_{{\bf k} \beta}^{*}$ is responsible for transferring 
electrons from the QD the normal ($\beta\!=\!N$) or 
superconducting ($S$) leads. 

We assume that the normal electrode is described by the 
Hamiltonian of noninteracting fermions $\hat{H}_{N} \!=\!  
\sum_{{\bf k},\sigma} \left( \varepsilon_{{\bf k}N}\!-\!\mu_{N} 
\right)  \hat{c}_{{\bf k} \sigma N}^{\dagger} \hat{c}_{{\bf k} 
\sigma N}$. For description of the superconducting lead 
we use a general expression
\begin{eqnarray}
\hat{H}_{S}=\sum_{{\bf k},\sigma} \left( \varepsilon_{{\bf k}S}
\!-\! \mu_{S} \right)  \hat{c}_{{\bf k} \sigma S }^{\dagger} 
\hat{c}_{{\bf k} \sigma S} + \hat{V}_{pairing} ,
\end{eqnarray}
where the two-body term $\hat{V}_{pairing}$ induces either 
the coherent (for $T\!<\!T_{c}$) or incoherent electron pairs
(above $T_{c}$).


Without specifying $\hat{V}_{pairing}$ nor restricting to 
any particular microscopic mechanism of superconductivity 
we use the basic BCS-type results. For the superconducting 
state (below the critical temperature $T_{c}$) we introduce, 
in a standard way, the following retarded Green's function 
expressed in the Nambu representation 
\begin{eqnarray}
G^{r}_{S}({\bf k},\omega)^{-1}  = \left( 
\begin{array}{cc}
\omega - \xi_{{\bf k}S} & \Delta_{\bf k} \\
\Delta_{\bf k} & \omega + \xi_{{\bf k}S} 
\end{array} \right)  
\label{retarded_GF}
\end{eqnarray}
where $\xi_{{\bf k}\beta}\!=\!\varepsilon_{{\bf k}\beta}-\mu
_{\beta}$ measures the energies from the chemical potential. The 
off-diagonal terms $\Delta_{\bf k}$ have as usually a meaning of 
the gap in the single particle excitation spectrum of $S$ electrons. 

Since the excitation gap is known to develop well above 
the transition temperature we follow the arguments of Levin 
et al \cite{Levin-05} and impose the following phenomenological 
Ansatz 
\begin{eqnarray}
\Delta_{\bf k}^{2} = \Delta_{{\bf k},sc}^{2} + 
\Delta_{{\bf k},pg}^{2} .
\label{two_gaps}
\end{eqnarray}
The first part in (\ref{two_gaps}) is related to the 
superconducting order parameter $\langle \hat{c}_{-{\bf k}
\downarrow S} \hat{c}_{{\bf k}\uparrow S} \rangle$ while 
the latter one comes from the pseudogap. For a quantitative 
discussion we propose the temperature dependence given by
\begin{eqnarray}
\Delta_{{\bf k},sc}(T) = \left\{ \begin{array}{ll}
\Delta_{\bf k}(0) \sqrt{1-\left(\frac{T}{T_{c}}\right)^{2}} 
&\hspace{0.2cm} \mbox{for} \hspace{0.2cm} 
T  \leq  T_{c} , \\ 0 &
\hspace{0.2cm} \mbox{for} \hspace{0.2cm}  
T > T_c  .  \end{array} \right.
\label{sc_gap}
\end{eqnarray}
The pseudogap contribution $\Delta_{{\bf k},pq}$ to the 
effective gap (\ref{two_gaps}) origins from the preformed 
electron pairs. Above $T_{c}$ their long range coherence 
is missing (hence a name of the incoherent pairs) and such 
pairs ultimately dissociate at a certain temperature $T^{*}$. 
In analogy to (\ref{sc_gap}) we propose to consider 
the following temperature dependence 
\begin{eqnarray}
\Delta_{\bf k}(T) = \Delta_{\bf k}(0) \;  
\sqrt{1-\left(\frac{T}{T^{*}}\right)^{2}} .
\label{pg_gap}
\end{eqnarray}
%

\begin{figure}
\centerline{\epsfxsize=6cm \epsffile{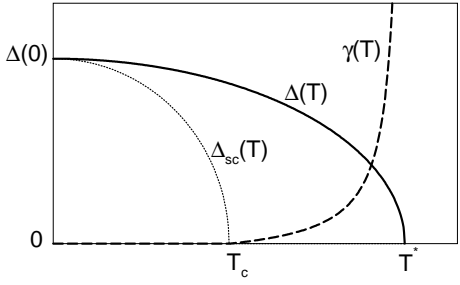}}
\caption{Temperature dependence of the single particle gap
$\Delta(T)$ (solid line), the damping rate $\gamma(T)$ 
(dashed line) and the superconducting order parameter
$\Delta_{sc}(T)$ (dotted line).}
\label{Fig1}
\end{figure}

Absence of the off-diagonal long range order above $T_{c}$ 
implies that the retarded Green's function (\ref{retarded_GF}) 
must reduce to a diagonal structure for all temperatures 
$T_{c}\!<\!T\!<\!T^{*}$. In such a case the pseudogap affects 
the single particle spectrum only through the BCS-like shape 
of the diagonal part \cite{Levin-05} 
\begin{eqnarray}
&& G^{r}_{S}({\bf k},\omega) = \label{G_pg} \\
&& \left( \begin{array}{cc} 
\frac{u_{\bf k}^{2}}{\omega - E_{\bf k} + i \gamma_{\bf k}}+
\frac{v_{\bf k}^{2}}{\omega + E_{\bf k} + i \gamma_{\bf k}}
 & 0 \\ 0 & 
\frac{v_{\bf k}^{2}}{\omega - E_{\bf k} + i \gamma_{\bf k}}+
\frac{u_{\bf k}^{2}}{\omega + E_{\bf k} + i \gamma_{\bf k}}
\end{array}\right)
\nonumber
\end{eqnarray}
with the quasiparticle dispersion  $E_{\bf k}=\sqrt{\xi_{{\bf k}S}^{2}
+\Delta_{\bf k}^{2}}$ and the usual coherence factors $u_{\bf k}^{2}=
\frac{1}{2}\left(1+\frac{\xi_{{\bf k}S}}{E_{\bf k}}\right)=1\!-
\!v_{\bf k}^{2}$. This sort of behavior (\ref{G_pg}) can be derived 
on a microscopic level investigating the pairing interactions beyond
the mean-field BCS framework \cite{beyond-BCS}.

In addition to (\ref{sc_gap}) and (\ref{pg_gap}) we have chosen 
for computational purposes some phenomenological damping rate
\begin{eqnarray}
\gamma_{\bf k} = \left\{ 
\begin{array}{ll}
0^{+} &\hspace{0.2cm} \mbox{for} \hspace{0.2cm} T  \leq  T_{c} ,\\
 \gamma_{\bf k}(0) \; \frac{T-T_{c}}{T^{*}-T}  & \hspace{0.2cm} 
\mbox{for} \hspace{0.2cm} T_c  <  T \! \leq \! T^{*} 
\end{array} \right.
\label{damping}
\end{eqnarray}
and assumed its momentum variation in the following way 
$\gamma_{\bf k}= \gamma^{2}/( \gamma+\frac{|\xi_{{\bf k}S}|}{1000})$ 
where parameter $\gamma \equiv \gamma_{{\bf k}_{F}}$. Figure (\ref{Fig1}) 
illustrates the temperature dependences described in this section.

\section{The uncorrelated QD}

To introduce the formalism of our calculations we first start
analyzing the equilibrium case $\mu_{L}\!=\!\mu_{R}$. In the 
standard Nambu notation we express the retarded Green's function 
of the QD through the Dyson equation
\begin{eqnarray}
G^{r}_{d}(\omega)^{-1} = g^{r}_{d}(\omega)^{-1} 
- \Sigma^{r}_{d}(\omega) 
\label{Dyson}
\end{eqnarray}
with two contributions to the matrix selfenergy $\Sigma^{r}_{d}
(\omega)=\Sigma^{r}_{N}(\omega)+\Sigma^{r}_{S}(\omega)$. This 
problem can be solved exactly in a case of the noninteracting 
QD ($U\!=\!0$) when 
\begin{eqnarray}
g^{0r}_{d}(\omega)^{-1} = 
\left( \begin{array}{cc} 
\omega - \varepsilon_{d} + i 0^{+} & 0\\
0 & \omega + \varepsilon_{d} + i 0^{+}
\end{array}\right)
\end{eqnarray}
and the selfenergies simplify to \cite{Krawiec-04}
\begin{eqnarray}
&& \Sigma^{0r}_{N}(\omega) =  
\left( \begin{array}{cc} 
\sum_{\bf k} \frac{\left| V_{{\bf k}N} \right|^{2}}
{\omega - \xi_{{\bf k}N} + i 0^{+}} & 0\\
0& \sum_{\bf k} \frac{\left| V_{{\bf k}N} \right|^{2}}
{\omega + \xi_{{\bf k}N} + i 0^{+}}
\end{array}\right)
\label{Sigma_N} \\
&& \Sigma^{0r}_{S}(\omega) = \sum_{\bf k} 
\left| V_{{\bf k}S} \right|^{2} \times \label{Sigma_S} \\
&& \left( \begin{array}{cc} 
\frac{u_{\bf k}^{2}}{\omega - E_{\bf k} + i \gamma_{\bf k}}+
\frac{v_{\bf k}^{2}}{\omega + E_{\bf k} + i \gamma_{\bf k}}& 
\frac{u_{\bf k}v_{\bf k}}{\omega + E_{\bf k} + i \gamma_{\bf k}}-
\frac{u_{\bf k}v_{\bf k}}{\omega - E_{\bf k} + i \gamma_{\bf k}}\\ 
\frac{u_{\bf k}v_{\bf k}}{\omega + E_{\bf k} + i \gamma_{\bf k}}-
\frac{u_{\bf k}v_{\bf k}}{\omega - E_{\bf k} + i \gamma_{\bf k}}& 
\frac{v_{\bf k}^{2}}{\omega - E_{\bf k} + i \gamma_{\bf k}}+
\frac{u_{\bf k}^{2}}{\omega + E_{\bf k} + i \gamma_{\bf k}}
\end{array}\right) .
\nonumber
\end{eqnarray}
The BCS coefficients $u_{\bf k} v_{\bf k} = \frac{\Delta_{{\bf k},
sc}}{2E_{\bf k}}$ assure that a proximity effect in the QD occurs 
only for temperatures $T\!<\!T_{c}$. Otherwise, the pseudogap 
appearing in the spectrum of $S$ electrons affects the Green's 
function (\ref{G_pg}) via the diagonal parts of (\ref{Sigma_S}).

\begin{figure}
\centerline{\epsfxsize=6.7cm \epsffile{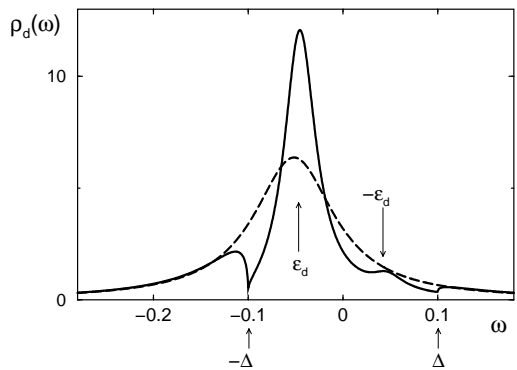}}
\caption{Spectral function $\rho_{d}(\omega)$ of the QD 
for $U\!=\!0$ with the right h.s.\ electrode 
being in a superconducting (solid line) and in a normal 
state (dashed line). We used the isotropic energy gap 
$\Delta_{\bf k}=0.1 D$ and $\varepsilon_{d}=-0.05D$, 
$\Gamma_{\beta}=0.05 D$, $\gamma=0.01D$ and set the 
half-bandwidth $D$ as a unit for energies.}
\label{Fig2}
\end{figure}

\begin{figure}
\centerline{\epsfxsize=6.5cm \epsffile{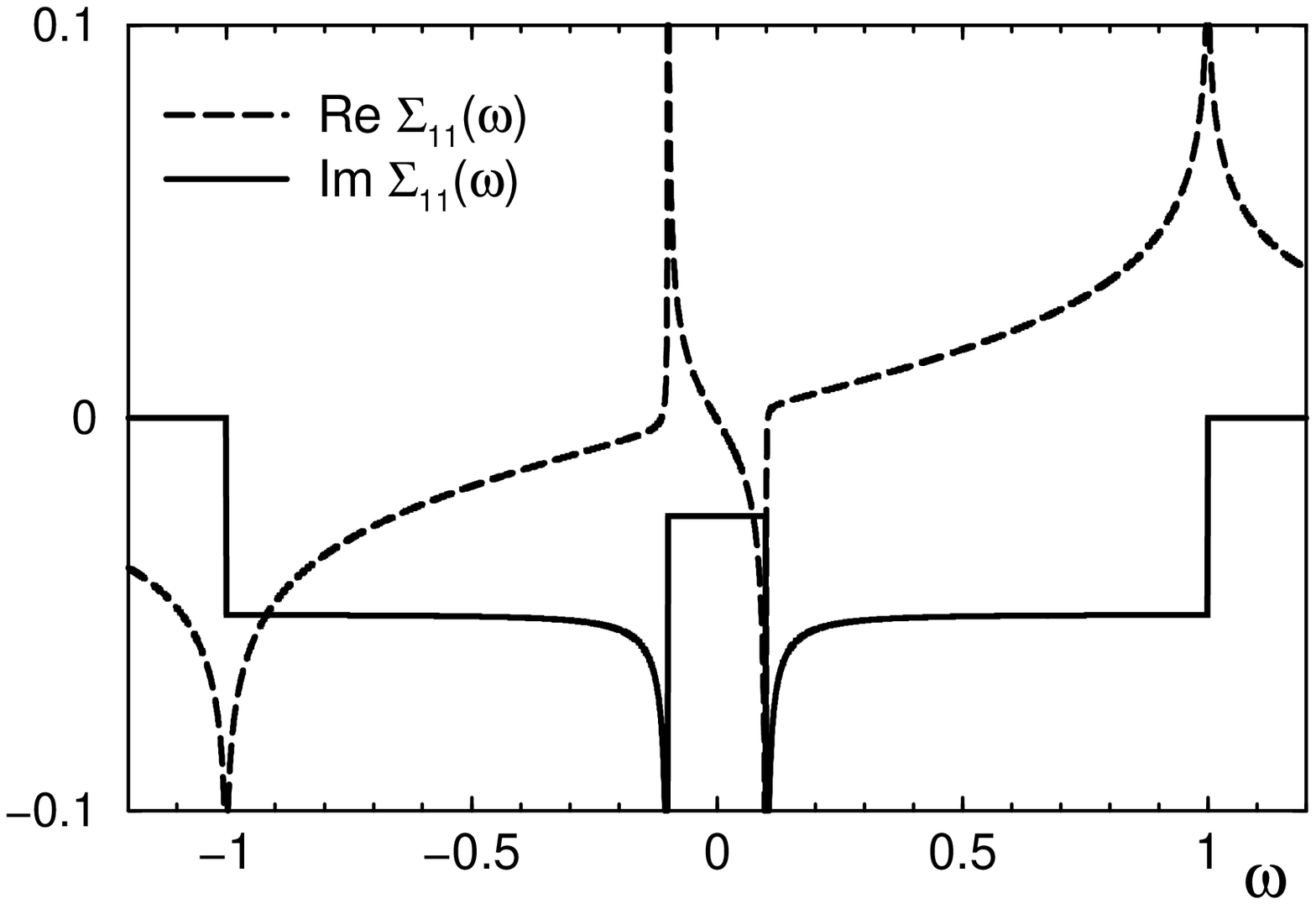}}
\centerline{\epsfxsize=6.5cm \epsffile{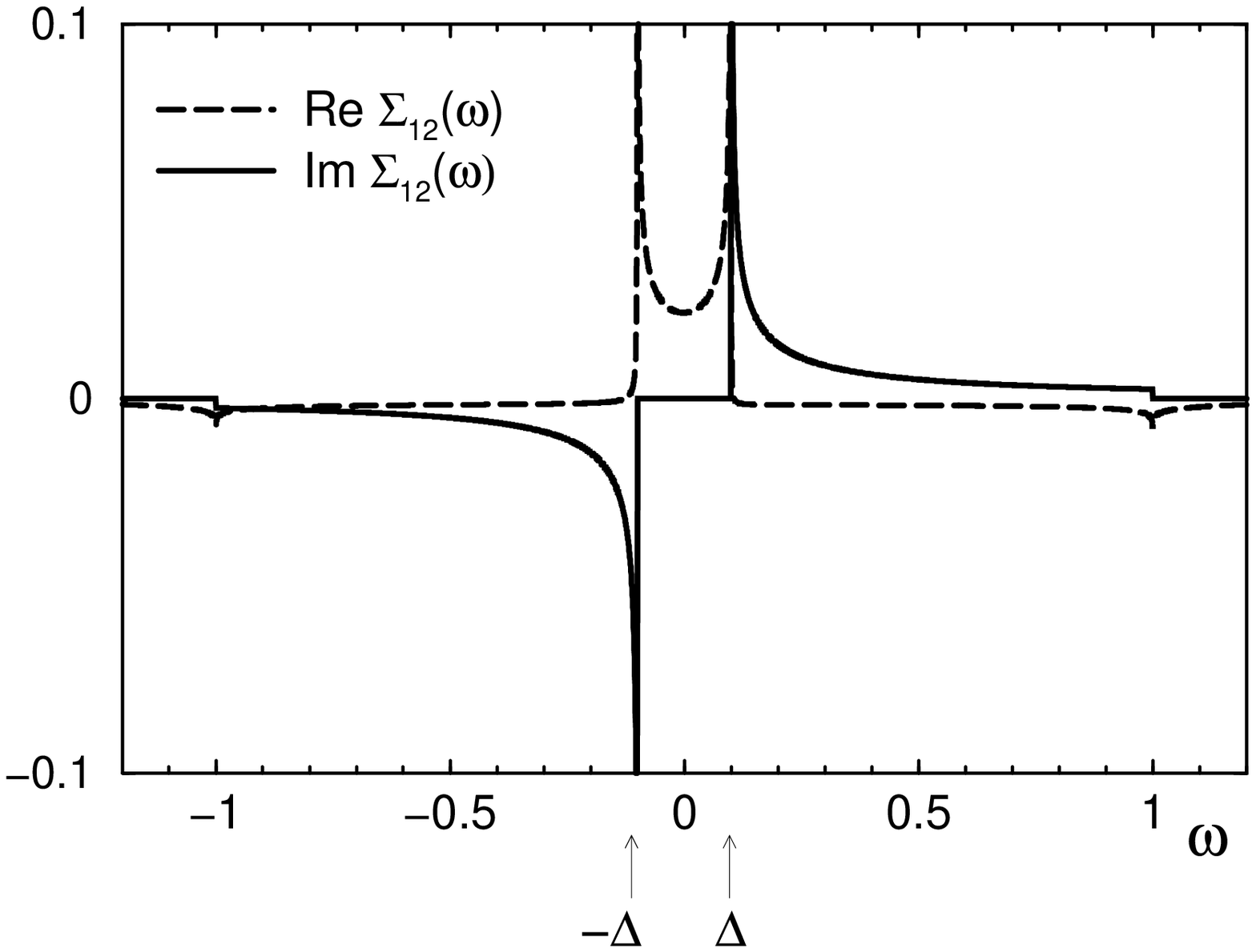}}
\caption{The real and imaginary parts of the matrix
selfenergy for the diagonal $\Sigma_{11}^{r}(\omega)$ 
(top panel) and off-diagonal terms $\Sigma_{12}^{r}
(\omega)$ (bottom panel). We used the same set of 
parameters as in figure \ref{Fig1}.}
\label{Fig3}
\end{figure}

The hybridization coupling $V_{{\bf k}\beta}$ can be conveniently 
replaced by the weighted density function $\Gamma_{\beta}(\omega)
\!=\! 2\pi \sum_{\bf k} |V_{{\bf k}\beta}|^{2} \delta(\omega\!-\!
\varepsilon_{{\bf k}\beta})$. For both electrodes being normal 
the QD spectral function $\rho_{d}(\omega)=-\frac{1}{\pi} G_{d
\;11}^{r}(\omega)$ acquires a Lorentzian shape centered around 
the single particle level $\varepsilon_{d}$ (see the dashed line 
in figure \ref{Fig2}) with the  effective broadening $\Gamma\!=\!
\Gamma_{N}\!+\!\Gamma_{S}\!=\!2\Gamma_{N}$. 

\begin{figure}
\centerline{\epsfxsize=9cm \epsffile{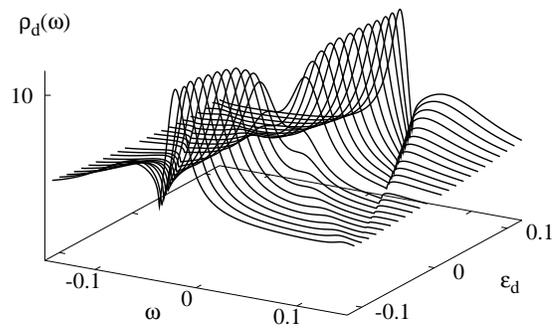}}
\caption{The ground state spectral function $\rho_{d}(\omega)$ 
of the QD for $U\!=\!0$ versus a varying position of the energy 
level $\varepsilon_{d}$. One can note a clear particle-hole 
mixing (two Lorentians built around $\pm \varepsilon_{d}$).}
\label{Fig_xxx}
\end{figure}

If one of the electrodes is a superconductor with an isotropic 
(${\bf k}$ independent) energy gap we can notice some qualitative 
(and quantitative) differences in the QD spectrum.
\begin{itemize}
\item[{(i)}] 
Since $S$ electrons can occupy no states in the energy gap 
$|\omega| < \Delta$ thereby the line broadening gets reduced by 50$\% 
$ and in consequence the QD peak around $\varepsilon_{d}$ 
becomes more pronounced.
\item[{(ii)}] 
A large amount of $S$ electron states is cumulated near $\omega 
= \pm \Delta$ (i.e.\ at the square root divergences in the density 
of the $S$ lead). Efficiency of the hybridization $V_{{\bf k}S}$ 
is there considerably enhanced leading to a depletion of the QD 
states at $\omega= \pm \Delta$.
\item[{(iii)}] 
A role of well defined quasiparticles in the superconducting
state is played by the electron pairs. Due to the hybridization 
$V_{{\bf k}S}$ such a particle hole mixing is also transfered 
onto the QD spectrum (the proximity effect). Indeed, in figures 
\ref{Fig2} and \ref{Fig_xxx} we notice that besides the Lorentzian 
peak centered around $\varepsilon_{d}$ there also appears its tiny 
mirror reflection at  $-\varepsilon_{d}$.
\end{itemize}

To provide the arguments for the above mentioned effects we 
plot in figure \ref{Fig3} the diagonal and off-diagonal parts 
of the matrix selfenergy $\Sigma_{d}(\omega)$. An odd symmetry 
of $\mbox{Im} \left[ \Sigma_{12}(\omega) \right]$ gives a 
non-vanishing $\mbox{Re} \left[ \Sigma_{12}(\omega)\right]$ for 
all energies located inside the energy gap $|\omega|<\Delta$. 
In diagonal term $\Sigma_{11}(\omega)$ the imaginary part is 
even (and negative) while the real part is odd (therefore
vanishing at $\omega\!=\!0$). Similar quantitative behavior 
of the matrix selfenergy $\Sigma_{d}(\omega)$ has been 
previously reported by several authors \cite{Fazio-98,Sun-99,
Clerk-00,Cuevas-01}, however, no clear evidence of the 
particle-hole mixing has been emphasized so far.

\section{The strong correlation limit}

It is well known both from the theoretical \cite{Meir-92} 
and experimental studies \cite{Goldhaber-98,Franceschi} that
the Coulomb interactions have a remarkable influence on 
transport properties through the QD. In particular, such
correlations are responsible for the Coulomb blockade 
(observed by oscillations of the differential conductance) 
and, at sufficiently low temperatures, produce the Kondo 
resonance leading to enhancement of the conductance 
to the unitary limit value $2e^2/h$. 

In this section we consider the correlations focusing on 
the extreme limit of $U\!=\!\infty$. Under such condition 
no double occupancy of the QD is allowed and one expects 
it to have a tremendous effect on the charge tunneling 
especially in the anomalous channels involving the 
electron pairs.

Excluding the doubly occupied states from the Hilbert space 
can be formally achieved using the auxiliary fields
\begin{eqnarray}
\hat{d}_{\sigma}^{\dagger} = \hat{f}_{\sigma}^{\dagger} \hat{b}
\hspace{0.5cm} 
\hat{d}_{\sigma}= \hat{b}^{\dagger} \hat{f}_{\sigma} 
\end{eqnarray}
where the boson $\hat{b}^{(\dagger)}$ and fermion operators 
$\hat{f}_{\sigma}^{(\dagger)}$ correspond to an annihilation 
(creation) of the empty and singly occupied states on the QD. 
These new fields must obey the local constraint $\hat{b}
^{\dagger}\hat{b} + \sum_{\sigma} \hat{f}_{\sigma}^{\dagger}
\hat{f}_{\sigma}=1$.

There are various methods to deal with the local constraint. 
For simplicity, we apply here the technique proposed by 
Le Guillou and Ragoucy \cite{LeGuillou} where projecting 
out of the doubly occupied states is achieved by appropriate
commutation relations between the operators of auxiliary 
fields. For the present context (\ref{model}) some necessary 
technical details have been previously discussed in Ref.\ 
\cite{Krawiec-04}. 

\begin{figure}
\centerline{\epsfxsize=7cm \epsffile{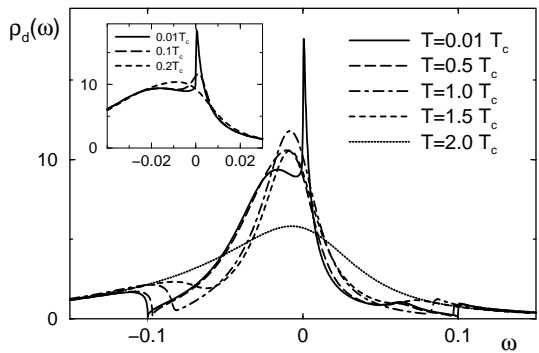}}
\caption{Spectral function $\rho_{d}(\omega)$ of the QD 
in the limit $U\!=\!\infty$ obtained for $\varepsilon_{d}
=-0.05D$, $\Gamma_{\beta}=0.01D$ assuming the isotropic
energy gap $\Delta_{\bf k}(T\!=\!0)=0.1D$ and $T^{*}=2T_{c}$.}
\label{Fig4}
\end{figure}

In the limit $U\!=\!\infty$ the Dyson equation (\ref{Dyson}) 
can be solved using the renormalized propagator $g^{r}_{d}
(\omega)=\left( 1 - n_{-\sigma} \right) g^{r0}_{d}(\omega)$ 
and the matrix selfenergy 
\begin{eqnarray}
\Sigma^{r}_{\beta}(\omega) = \left[ \Sigma^{0r}_{\beta}
(\omega) + \Sigma^{Ir}_{\beta}(\omega) \right] / 
\left( 1 - n_{-\sigma} \right)
\label{Sigma_infty}
\end{eqnarray}
where the contribution $\Sigma^{0r}_{\beta}$ of 
noninteracting electrons is given in (\ref{Sigma_N},
\ref{Sigma_S}). The other contribution $\Sigma^{Ir}_{\beta}$ 
originates from the correlations and under appropriate conditions 
leads to the Kondo effect \cite{Meir-92}. One finds \cite{Krawiec-04} 
\begin{eqnarray}
\Sigma^{Ir}_{\beta}(\omega) =  
n_{{\bf k}\beta} \; \tau_{3} \; \Sigma^{0r}_{\beta}
(\omega) \; \tau_{3}
\end{eqnarray}
where $\tau_3$ is the Pauli matrix and $n_{{\bf k}\beta}$
denotes an average occupancy of the ${\bf k}$-momentum in 
the $\beta$-th lead given by 
\begin{eqnarray}
n_{{\bf k}\beta}  = \left\{ \begin{array}{ll}
\left[ 1+ \mbox{exp}(\frac{\xi_{{\bf k}N}}{k_{B}T}) \right]^{-1} 
& \hspace{0.2cm} \mbox{for } \beta = N , \\ 
\frac{1}{2} \left[ 1 - \frac{\xi_{{\bf k}S}}{E_{\bf k}}
\mbox{tanh}\left( \frac{E_{\bf k}}{2k_{B}T} \right) \right]
&\hspace{0.2cm} \mbox{for } \beta=S .
  \end{array} \right.
\label{n_k}
\end{eqnarray}

In figure \ref{Fig4} we show the spectral function 
$\rho_{d}(\omega)$ calculated for several temperatures
in both, the superconducting and pseudogap states. In 
a comparison to the previous situation $U\!=\!0$ we can 
notice that:
\begin{itemize}
\item[{(i)}] for temperatures $T\!<\!T_{c}$ there are 
visible two Lorentzian peaks, however their positions are 
more shifted from $\pm \varepsilon_{d}$ because of a finite 
real part of the matrix selfenergy (\ref{Sigma_infty}),
\item[{(ii)}] 
for very low temperatures ($T<T_{K}$) there appears a narrow 
Kondo resonance at the Fermi energy associated with the spin 
singlet made of the QD and itinerant electrons (see the inset 
in Fig.\ \ref{Fig4}),
\item[{(iii)}] 
in the pseudogap regime above $T_{c}$ we no longer observe 
a tiny Lorentzian at $\omega \simeq - \varepsilon_{d}$ and 
simultaneously a dip of the spectral function at $| \omega| 
= \pm \Delta_{pg}(T)$ gets smeared because of the damping 
effects.
 \end{itemize}
For the particular set of parameters $\Gamma_{\beta}=0.01D$, 
$\varepsilon_{d}$ used in figure \ref{Fig4} we estimate the 
Kondo peak disappears for temperatures higher than $T_{K} 
\simeq 0.15 T_{c}$.On the other hand, for temperature exceeding 
$T^{*}$ the QD spectrum evolves back to its single Lorentzian 
peak centered around $\varepsilon_{d}$.

\section{Transport properties}

In order to study the nonequilibrium physics we use the 
Keldysh formalism. Applying a bias $V$ leads to imbalance 
of the chemical potentials $\mu_{N}\!-\!\mu_{S}\!=\!eV$ 
which induces the charge current $J(V)=-e\frac{d}{dt} 
\sum_{{\bf k},\sigma} \langle c_{{\bf k} N\sigma}^{\dagger} 
c_{{\bf k}N\sigma} \rangle$ through the QD.  

\begin{figure}
\centerline{\epsfxsize=7cm \epsffile{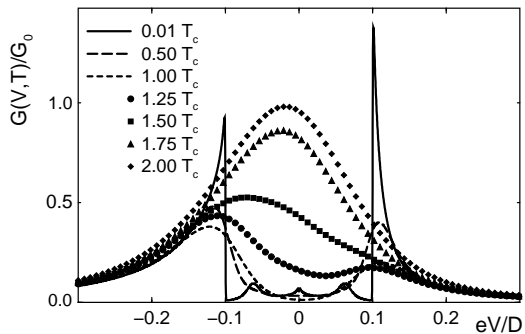}}
\caption{The differential conductance $G(V,T)$ versus the 
applied bias $V$ for a representative set of temperatures 
in the superconducting region (the lines) and for the
pseudogap phase (the symbols). We used the same set 
of parameters as in figure \ref{Fig4} and conductnce
is expressed in units of $G_{0}\equiv G(0,T^{*})$.}
\label{cond_vs_V}
\end{figure}

\begin{figure}
\centerline{\epsfxsize=7cm \epsffile{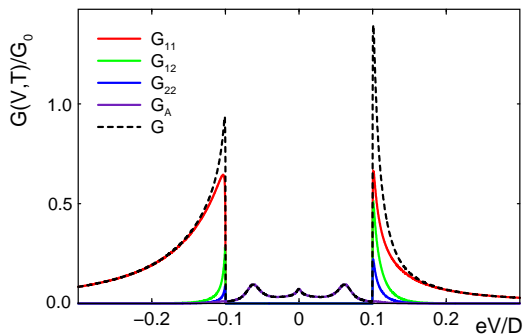}}
\caption{Contributions of the normal $G_{11}$ (the red line) 
and the anomalous channels $G_{12}$ (blue), $G_{22}$ (green), 
$G_{A}$ (indigo) to the total differential conductance $G(V,T)$. 
Temperature is $T=0.01 T_{c}$ (i.e.\ $T<T_{K}$) and other
parameters have the same values as in figure \ref{Fig4}.}
\label{cond_contrib}
\end{figure}

Following the procedure described previously \cite{Krawiec-04} 
we express the charge current $J(V)$ in terms of the following 
contributions 
\begin{eqnarray}
J=J_{11}+J_{12}+J_{22}+J_{A}.
\label{current}
\end{eqnarray}
The first three components in equation (\ref{current}) 
have the Landauer-type structure
\begin{eqnarray}
J_{ij}(V) = \frac{2e}{h} \int d\omega \;
T_{ij}(V) \left[ f(\omega-eV)-f(\omega) \right]
\label{J_ij}
\end{eqnarray}
where the corresponding transmittance are defined by
\begin{eqnarray}
T_{11}(V) & = & - \mbox{Im}\Sigma_{11,S}^{r} \; 
\left| G_{11}\right|^{2} \; \Gamma_{N}(\omega) ,
\label{T_11}\\
T_{12}(V) & = & - 2 \mbox{Im}\Sigma_{12,S}^{r} \; 
\mbox{Re} \left[G_{11} G^{*}_{12} \right] \; 
\Gamma_{N}(\omega) , \\
T_{22}(V) & = & - \mbox{Im}\Sigma_{22,S}^{r} \; 
\left| G_{12}\right|^{2} \Gamma_{N}(\omega) .
\end{eqnarray}
The last contribution describes the Andreev current 
\begin{eqnarray}
J_{A}(V) = \frac{2e}{h} \int d\omega \;
T_{A}(V) \left[ f(\omega-eV)-f(\omega+eV) \right]
\label{J_A}
\end{eqnarray} 
with
\begin{eqnarray}
T_{A}(V)  = - \mbox{Im}\Sigma_{22,N}^{r} \; 
\left| G_{12}\right|^{2} \Gamma_{N}(\omega) .
\end{eqnarray}
This kind of current arises when electron from the $N$ lead 
is converted into the Copper pair in the $S$ electrode and 
simultaneously a hole is reflected back to the $N$ lead. For 
a detailed discussion of such anomalous Andreev current
see for instance the recent review article \cite{Deutscher-05}.
In a case when the both leads are normal (i.e.\ $\Delta 
\rightarrow 0$) there survives only $J_{11}(V)$ current 
and moreover the transmittance (\ref{T_11}) simplifies to 
the usual form $T(\omega)=\rho_{d}(\omega) \; \frac{
\Gamma_{N}(\omega)\Gamma_{s}(\omega)}{\Gamma_{N}(\omega)
+\Gamma_{S}(\omega)}$ \cite{Meir-92}.

In figure \ref{cond_vs_V} we plot the differential conductance
$G(V,T)=d J(V)/dV$ as a function of the external bias $V$ for 
several representative temperatures. In the superconducting state
we clearly notice a strong suppression of the charge current at 
small voltages $|eV| \leq \Delta(T)$. Due to energy gap in the
spectrum of $S$ electrons the only possible process of a charge 
tunneling occurs then through the Andreev channel. At temperatures 
below $T_{c}$ the non-vanishing conductance $G(V,T) \simeq G_{A}
(V,T)$ for $|eV| \leq \Delta(T)$ is almost an order of 
magnitude smaller than the normal state conductance $G(V,T^{*})$.

\begin{figure}
\centerline{\epsfxsize=7cm \epsffile{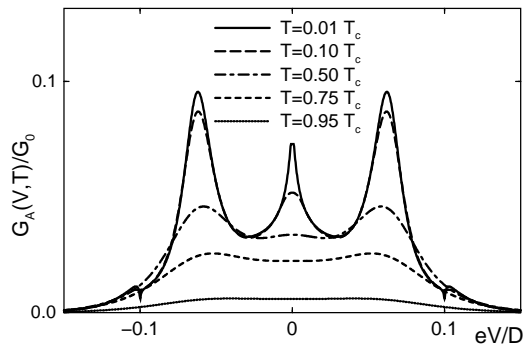}}
\caption{The differential conductance $G_{A}(V,T)$ of the 
Andreev current for a number of temperatures (see the legend). 
The zero bias enhancement (central peak) is due to the Kondo 
resonance. Above  $T_{c}$ the Andreev 
current vanishes \cite{Deutscher-05}.}
\label{condA_vs_V}
\end{figure}

\begin{figure}
\centerline{\epsfxsize=6cm \epsffile{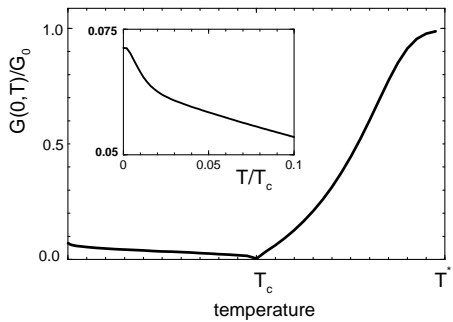}}
\caption{Temperature dependence of the zero bias differential 
conductance $G(V\!=\!0,T)$ for the set of parameters used 
in figure \ref{Fig4}. Inset shows the enhancement at low 
temperatures which is due to the Kondo resonance.}
\label{cond_vs_T}
\end{figure}

For sufficiently low temperatures $T\!<\!T_{K}$, we observe 
formation of the Kondo peak (see figure \ref{Fig4}) which affects
the differential conductance at small bias $|V|$. In the present 
it gives some enhancement of the zero bias conductance (see the 
inset of figure \ref{cond_vs_T}) through the Andreev scattering. 
This zero bias anomaly is however rather residual as compared to 
the N-QD-N junctions \cite{Meir-92}. A strong suppression of the 
zero bias anomaly has been previously pointed out by several 
authors \cite{Fazio-98}. We checked that the low temperature 
differential conductance $G(0,T)$ fits very well the universal 
parabolic variation characteristic for the Kondo regime.

For higher voltages, exceeding $\Delta(T)$, the dominant role 
to the charge transport comes from the normal current $J_{11}(V)$.
In figure \ref{cond_contrib} we show each of the contributions to 
the total conductance for a small temperature $T\!<\!T_K\!<\!T_c$. 
The anomalous channels $J_{12}(V)$ and $J_{22}(V)$ get activated 
outside the energy gap. But these contributions, as well as 
$G_{A}(V)$, do quickly diminish for an increasing bias $|V|$.

The in-gap conductance arising from the Andreev current is 
very sensitive to the temperature (see figure \ref{condA_vs_V}).
Already for temperatures higher then $T \sim T_{K}$ the Kondo 
peak is washed out which gives a concomitant disappearance of 
the zero bias anomaly. Upon further increasing the temperature 
there occurs a gradual suppression of the Andreev current 
which completely vanishes at $T \rightarrow T_{c}^{-}$. 

For temperatures above $T_{c}$ the charge current is transmitted 
only via the normal $J_{11}(V)$ channel. With an increasing temperature  
the pseudogap becomes gradually filled in (see figure \ref{cond_vs_V}) 
therefore the zero bias conductance smoothly increases, reaching  
a local maximum at $T^{*}$. From basic considerations \cite{Meir-92} 
it is well known that in the normal state (for $T\!>\!T^{*}$) the 
differential conductance exponentially decreases with respect to 
$T$. We would like to stress that the non-monotonous temperature 
variation of the zero bias differential conductance $G(0,T)$ 
(shown in figure \ref{cond_vs_T}) could be a useful method for 
identifying the temperature region of the incoherent electron 
pairs.

\section{Conclusions}

We have investigated the charge tunneling from the normal (N) to
superconducting (S) leads via the quantum dot (QD). In agreement
with the previous studies \cite{Fazio-98,Sun-99,Clerk-00,Cuevas-01}
we find that the long range off-diagonal superconducting order 
induces the proximity effect in the quantum dot (independently 
of the on-dot Coulomb interaction $U$). Besides the main 
Lorentzian peak near the single particle energy level $\varepsilon_{d}$ 
there appears another small counter-peak around $-\varepsilon_{d}$ 
due to the particle-hole mixing. Such physical situation occurs 
for all temperatures below $T_{c}$. This effect could be 
practically observed in measurements of the differential 
conductance upon varying the gate voltage.

In a presence of the strong on-dot correlations $U$ there arise 
some additional qualitative features. One of them is the Kondo 
state which occurs at sufficiently low temperatures, usually 
below 1K \cite{Goldhaber-98}. Spectral function of the QD develops 
then a narrow peak at the Fermi energy. It has a qualitative 
influence on the low energy transport properties. Since the 
normal channel $J_{11}(V)$ is forbidden for $|eV| \!<\Delta$ 
(because of the energy gap of $S$ electrons) the only 
possibility for the charge transfer is through the Andreev 
tunneling. A magnitude of the Andreev current is by far 
smaller than the normal charge current but nevertheless, 
in analogy to the N-QD-N Kondo anomaly \cite{Meir-92,Goldhaber-98}, 
we find some enhancement of the zero bias conductance for the 
N-QD-S setup. This anomaly shows up at temperatures $T<T_{K}$ 
but the conductance does not reach the unitary limit value 
$2e^{2}/h$.

We have also extended our study on the pseudogap state and found
that incoherent electron pairs have a remarkable influence on the 
charge transport. Above $T_{c}$ the zero bias differential conductance 
$G(0,T)$ systematically increases upon increasing temperature because 
the pseudogap is gradually filled in. Finally, at $T^*$, it reaches 
a local maximum and for higher temperatures (in the normal state) 
has a changeover to the exponential decrease versus $T$. Such 
nonmonotonus variation of the zero bias conductance (figure 
\ref{cond_vs_T}) can serve as a sensitive method to identify 
the temperature region $T_c<T<T^*$ of the incoherent electron 
pairs. 

Some other effects which might eventually come into the play. One 
of them is a strong anisotropy of the gap function $\Delta_{\bf k}$ 
well known for all the HTSC cuprates. To some extent this issue 
has been recently addressed by ETH group \cite{Sigrist-06}. Authors 
investigated the conductance of charge current flowing between 
the conducting STM tip and the quasi-2D superconducting CuO$_{2}$ 
planes involving the apical oxygen atoms (to be regarded as the QD).
They focused on studying the inelastic scattering caused by the 
oxygen atoms' vibrations. In the future work we shall discuss 
the normal and anomalous currents for such anisotropic gap 
$\Delta_{\bf k}$ of the $d$-wave symmetry. As known from the
literature the gapless superconductivity does not preclude 
appearance of the Kondo resonance \cite{Borkowski-92,Kondo_in_dwave}.

\section{Acknowledgments }

This work is partially supported by Polish Committee 
of Scientific Research under the grant No.\ 2P03B06225.

\end{document}